\documentstyle[11pt,epsfig,axodraw]{article}
 \def\baselinestretch{1.3}
\newcommand{\ba}{\begin{array}}
\newcommand{\ea}{\end{array}}
\newcommand{\bd}{\begin{displaymath}}
\newcommand{\ed}{\end{displaymath}}
\newcommand{\be}{\begin{equation}}
\newcommand{\ee}{\end{equation}}
\newcommand{\bea}{\begin{eqnarray}}
\newcommand{\eea}{\end{eqnarray}}
\newcommand{\Dir}{\kern -6.4pt\Big{/}}
\newcommand{\Dirin}{\kern -10.4pt\Big{/}\kern 4.4pt}
\newcommand{\DDir}{\kern -10.6pt\Big{/}}
\newcommand{\DGir}{\kern -6.0pt\Big{/}}
\begin{document}
\setcounter{page}{0}
\thispagestyle{empty}
\
\def\bra{\langle}
\def\ket{\rangle}

\def\a{\alpha}
\def\as {\alpha_s}
\def\b{\beta}
\def\d{\delta}
\def\e{\epsilon}
\def\ve{\varepsilon}
\def\l{\lambda}
\def\m{\mu}
\def\n{\nu}
\def\G{\Gamma}
\def\D{\Delta}
\def\L{\Lambda}
\def\s{\sigma}
\def\p{\pi}

\def\etal{ {\em et al.}}
\def\mzs {M_Z^2}
\def\mws {M_W^2}
\def\q2 {q^2}
\def\sz {\sin^2\theta_W}
\def\cz {\cos^2\theta_W}
\def\lp{\lambda^{\prime}}
\def\lps{\lambda^{\prime *}}
\def\lpp{\lambda^{\prime\prime}}
\def\lpps{\lambda^{\prime\prime * }}
\def\lms{\Lambda^{(n_{\rm f}=4)}_{\overline{\mathrm{MS}}}}
\def\MSbar{\overline{\mathrm{MS}}}

\def\bapp{b_1^{\prime\prime}}
\def\bbpp{b_2^{\prime\prime}}
\def\bcp{b_3^{\prime}}
\def\bdp{b_4^{\prime}}
\def\t {\times }
\def\slash {\!\!\!\!\!\!/}
\def\photino {\tilde\gamma}
\def\sel {\tilde{e}}
 \def\N10{\widetilde \chi_1^0}
                         \def\C1p{\widetilde \chi_1^+}
                         \def\C1m{\widetilde \chi_1^-}
                         \def\C1pm{\widetilde \chi_1^\pm}
 \def\Ntwo{\widetilde \chi_2^0}
                         \def\Ctwo{\widetilde \chi_2^\pm}
\def\lslep {{\tilde e}_L}
\def\rslep {{\tilde e}_R}
\def\sneu {\tilde \nu}
\def\msneu {M_\tilde \nu}
\def\mrslep {m_{\rslep}}
\def\mlslep {m_{\lslep}}
\def\mneu {m_{\neu}}
\def\mpT{p_T \hspace{-1em}/\;\:}
\def\mET{E_T \hspace{-1.1em}/\;\:}
\def\mE{E \hspace{-.7em}/\;\:}
\def\go{\rightarrow}
\def\beq{\begin{eqnarray}}
\def\Rp{R\!\!\!\!/}
\def\wrp {{\cal W}_{R\!\!\!\!/}}
\def\enq{\end{eqnarray}}
\def\goes{\longrightarrow}
\def\lsim{\:\raisebox{-0.5ex}{$\stackrel{\textstyle<}{\sim}$}\:}
\def\gsim{\:\raisebox{-0.5ex}{$\stackrel{\textstyle>}{\sim}$}\:}
 \begin{flushright}
{\large DFTT 15/03}\\ 
{\large SHEP-03-11}\\
{\large July 2003}
\end{flushright}
\begin{flushleft}
\end{flushleft}
\begin{center}
{\Large\bf One-loop weak corrections to the $b\bar b$ cross section}\\
{\Large\bf at TeV energy hadron colliders}\\[10mm]
{\Large E. Maina$^{a,}$\footnote{maina@to.infn.it},
        S. Moretti$^{b,}$\footnote{stefano@hep.phys.soton.ac.uk},
        M.R. Nolten$^{b,}$\footnote{mrn@hep.phys.soton.ac.uk} and
        D.A. Ross$^{b,}$\footnote{dar@hep.phys.soton.ac.uk}}\\[4mm]
{\em a) Dipartimento di Fisica Teorica and Sezione INFN,
Universit\`a di Torino,\\
Via Pietro Giuria 1, 10125 Torino, Italy}\\[5mm]
{\em 2) School of Physics \& Astronomy, University of Southampton,\\
Highfield, Southampton SO17 1BJ, UK}
\\[15mm]

\end{center}

\begin{abstract}
\noindent\small   
We investigate one-loop weak corrections  to the production cross section
of two $b$-jets  at Tevatron and Large Hadron Collider
(LHC). We establish that they
are small at inclusive level but dominant in exclusive observables that have
a non-trivial dependence on the helicity structure of the hard
subprocesses. Such  effects can serve as a test of the Standard Model (SM)
and, conversely, they should be taken into account in future 
experimental analyses aiming at extracting possible signals 
of new physics.

\end{abstract}

\vspace*{2.0truecm}

\clearpage

\section*{Introduction}

It has already been clearly established 
\cite{Kuroda:1991wn}--\cite{Melles:2001dh} that large 
Sudakov logarithms arising at TeV energy scales as a consequence of
a non-cancellation between real and virtual contributions can 
enhance the effects of Electro-Weak (EW) corrections in electron-positron
scattering, so that the latter grow
as $\alpha^n_{\mathrm{EW }}\log^{2n}(s/M_W^2)$ at the $n$-th perturbative
order even in fully inclusive observables, where
 $s_{e^+e^-}$ is the collider centre-of-mass (CM) energy
squared and $M_W$ the $W$ boson mass. Eventually, they can even surpass
the corrections generated in QCD: e.g., in the 
total hadronic cross section at
$\sqrt s_{e^+e^-}\approx800$ GeV and above.

The reason for this is intimately related to the 
violation of the Bloch-Nordsieck theorem occurring in non-Abelian theories
whenever the initial state has a finite (weak) isospin 
charge\footnote{The problem is in principle present 
also in QCD, with respect to the colour charge; in practice, however, 
it has no observable consequences, because of the final averaging of the 
colour degrees of freedom of the incoming partons, forced by their confinement
into colourless hadrons.},
as dictated by the given beam configuration.  
This is immediately evident for leptonic colliders, as the Sudakov logarithms
present in $e^+e^-$ scattering
would cancel against those originating in $e^+\nu_e$ and $\bar\nu_e e^-$
collisions (the (anti)neutrinos are the isospin partners of the
electron/positrons), a condition which is clearly  impossible to
satisfy experimentally. One can view the mechanism rather
intuitively from a diagrammatic perspective. In short, virtual $W$
corrections simply multiply the leading-order (LO) 
scattering matrix elements, thus being  proportional to $\sigma_{e^+e^-}$,
while the real emission of a $W$ boson does change the isospin of
the incoming electron/positron and turns it into a(n) (anti)neutrino,
so that the corrections here are proportional to 
 $\sigma_{e^+\nu_e}$ and $\sigma_{\bar\nu_e e^-}$.

Evidently, this does not occur for the case of real and virtual $Z$ boson 
corrections (or photons, for that matters). 
The source of the large logarithms is then in principle manifest only in 
the case of $W$ boson corrections. In practice, though, one should recall
that both $W$ and $Z$ real bosons are unstable and decay into high
transverse momentum leptons and/or jets, which are normally
captured by the detectors. In the definition of an
exclusive cross section,
one may then remove events with such additional particles.
Ultimately, other than being a second source of Bloch-Nordsieck
violation for the case of $W$ corrections,
this merely experimental
procedure will also spoil the cancellations between real and
virtual contributions  in the case of $Z$
bosons, simply because the former are not included in the
definition of the measured quantity. 

The leading, double--logarithmic, angular--independent weak
logarithmic corrections are universal, i.e. they depend only on the identities
of the external particles. Both leading and subleading corrections are finite
(unlike in
QED and QCD), as the masses of the $W$ and $Z$ gauge bosons provide a physical
cut-off for the otherwise divergent infrared behaviour.
In some instances large cancellation between angular--independent
and angular--dependent corrections \cite{Accomando:2001fn} and between
leading and
subleading corrections \cite{Kuhn:2001hz} have been found at TeV energies.
It is therefore of paramount importance to study the full set of fixed order
weak corrections in order to establish the relative size of the different
contributions at the energy scales which will be probed at TeV scale
machines.

Furthermore,  weak contributions can  be
isolated in a gauge-invariant manner from purely Electro-Magnetic
(EM) (or QED) effects
\cite{Ciafaloni:1999xg}, \cite{Beccaria:2000fk}--\cite{Beccaria:2001yf}, 
which may or may not
be included in the calculation, depending on the observable being studied
and the aimed at accuracy. In view of all such arguments,
it is then legitimate and topical to investigate
 the importance of higher-order weak effects at TeV 
scale hadronic colliders \cite{Accomando:2001fn}, 
such as Tevatron ($\sqrt s_{p\bar p}=2$ TeV)
and LHC ($\sqrt s_{pp}=14$ TeV). 

Some further considerations are however in order in the hadronic
context. First, one should recall that hadron-hadron scatterings 
($pp, p\bar p$)
involve valence (or sea) partons of opposite isospin in the same process.
Thus the above-mentioned cancellations may
{potentially} be restored. For example, in $p\bar p(pp)$ scatterings
one finds both $u\bar u(uu)$ and $u\bar d(ud)$ subprocess contributions
to the total hadronic cross section, which tend to balance each other,
this effect being actually 
modulated by the Parton Distribution Functions (PDFs).
Secondly, several crossing symmetries among the involved partonic subprocesses
can also easily lead to more cancellations. Thirdly, whether or not 
these two mechanisms take place, 
spin asymmetries due to weak effects would always be 
manifest in some observables, since 
QCD has a trivial helicity structure (just like QED).

The purpose of this paper is that of establishing the importance of one-loop
weak effects in $b$-jet production at Tevatron and LHC. This is a 
pressing problem, as the $p_T$ distribution of Tevatron data for
$b$-quark production shows a clear disagreement with the theory \cite{bTev},
now known to next-to-leading order (NLO) accuracy in QCD
\cite{NLO}\footnote{Also the
subleading LO tree-level contributions from EW interactions
have been calculated.}, 
even after all uncertainties related to the
definition of the cross section \cite{Field} and the
 extraction of the $b$-quark
fragmentation function are properly taken into account \cite{Cacciari}.
In order to avoid such uncertainties, we consider in this paper
the cross section for di-jet production for  which each jet contains
a $b$($\bar{b}$)-quark. Data from Run 2 
 is also expected to be presented
in this format \cite{bTev}. Comparisons of such $b$-jet cross sections
from Run 1 with NLO QCD \cite{Frixione}  show a less severe discrepancy
than in the case of $b$-quark distributions.
The comparison between theory and $b$-quark/jet 
data is eventually expected to continue at LHC 
with much higher precision \cite{bLHC}.

\section*{Production of $b$-jets  at Tevatron and LHC}

Even if the discrepancy referred to at the end of the previous section
may  not appear alarming at this stage,
it is conceivable that the higher statistics available after Run 2 
 will afford the possibility of looking at more exclusive
observables, in order to understand whether the difference may be due to
some possible new physics effects, such as, e.g., 
$W^\prime$ and $Z^\prime$ gauge bosons \cite{Baur}. In this respect, it is natural to
turn to quantities which are insensitive to QCD effects, such as the
aforementioned spin
induced asymmetries in the cross section.  From this point of view, the
knowledge of the weak effects described above would be of paramount
importance, even if their overall contribution to the inclusive cross
section should turn out to be negligible.

After Run 2 at Tevatron, the accumulated statistics will be sufficient 
to select hadronic samples with two $b$-jets and to establish
their charge as well: e.g., by extracting two displaced
vertices and measuring the charge of one of the (at least two) associated
jets, via a high $p_T$ lepton selection
or jet charge reconstruction. This will
enable one to define the usual `forward-backward asymmetry' for
$b$-jets also at hadronic colliders,  hereafter denoted by
$A_{\rm{FB}}$\footnote{In this respect, it is 
intriguing to recall   the long-standing
disagreement between data and SM for such a quantity,
as seen at LEP and  SLD \cite{Abb}, as well as
several other observables involving $b$-quarks/jets, both at collider
and fixed target experiments \cite{Stefano}.}. 
Unfortunately, because of the symmetric beam configuration at
LHC, one cannot define the forward-backward asymmetry in this case. 
Pure QCD contributions 
through orders $\alpha_{\mathrm{S}}^2$ and $\alpha_{\mathrm{S}}^3$
to such a quantity are negligible
at Tevatron compared to the tree-level EW ones,
which are of order $\alpha_{\mathrm{EW}}^2$. 
We set out here to compute
one-loop virtual effects to $b\bar b$ production through order
$\alpha_{\mathrm{S}}^2\alpha_{\mathrm{EW}}$, which have then formally
a similar strength to the purely EW ones, given that
$\alpha_{\mathrm{S}}^2\sim\alpha_{\mathrm{EW}}$ at TeV energies.

Before proceeding, we
should like to clarify here that we will only include (in 
the language of Ref.~\cite{Field}) `flavour creation' contributions and
neglect both the `flavour excitation' and `shower/frag\-men\-ta\-tion' ones.
While this is certainly not justified in the total inclusive $b$-cross 
section \cite{Field}, it is entirely appropriate for the $b\bar b$ one
that we will be using in the definition of $A_{\rm{FB}}$, for which we
will require `two' high $p_T$ $b$-jets
(thus depleting the `flavour excitation' terms) 
tagged in opposite hemispheres 
(thus suppressing  the `shower/fragmentation' contributions). 
Finally, as anticipated in the previous discussion, we will neglect
including QED corrections at this stage of our computation (this
is indeed a gauge-invariant procedure, as we have explicitly verified), 
since we will ultimately be most interested in the forward-backward asymmetry,
to which pure EM terms contribute negligibly.

\section*{Partonic contributions to the $pp/p\bar p\to b\bar b$ cross section}

The inclusive $b$-jet cross section at both Tevatron and LHC is dominated
by the pure QCD contributions
$gg\to b\bar b$ and $q\bar q\to b\bar b$, 
known through order $\alpha_{\rm S}^n$ for $n=2,3$.
Of particular relevance in this context is the fact
that for the flavour creation mechanisms no 
$\alpha_{\rm S}\alpha_{\rm{W}}$ tree-level contributions are allowed,
because of colour conservation: i.e., 
\vspace*{0.75truecm}
\begin{eqnarray}
\begin{picture}(300,-300)
\SetScale{1.0}
\SetWidth{1.2}
\SetOffset(0,-60)
\ArrowLine(30,90)(45,75)
\ArrowLine(45,75)(30,60)
\Gluon(45,75)(60,75){3}{3}
\Text(25,95)[]{\small $q$}
\Text(25,55)[]{\small $\bar q$}
\ArrowLine(60,75)(75,90)
\ArrowLine(75,60)(60,75)
\Text(80,95)[]{\small $b$}
\Text(80,55)[]{\small $\bar b$}
\Text(105,75)[]{$*~[$}
\ArrowLine(130,90)(145,75)
\ArrowLine(145,75)(130,60)
\Photon(145,75)(160,75){3}{3}
\Text(125,95)[]{\small $q$}
\Text(125,55)[]{\small $\bar q$}
\ArrowLine(160,75)(175,90)
\ArrowLine(175,60)(160,75)
\Text(180,95)[]{\small $b$}
\Text(180,55)[]{\small $\bar b$}
\Text(215.,75)[]{$]^\dagger~~=~~0$,}
\end{picture} 
\end{eqnarray}
\noindent
where the wavy line represents a $Z$ boson (or a photon) and the 
helical one a gluon. Tree-level asymmetric terms through 
the order $\alpha_{\rm{EW}}^2$ are however finite, as they
are given by non-zero quark-antiquark initiated
diagrams such as the one above wherein the gluon is
replaced by a $Z$ boson (or a photon).  The latter are the
leading contribution to the forward-backward asymmetry (more precisely,
those graphs containing one or two $Z$ bosons are, as those involving
two photons are subleading in this case, even with respect to
the pure QCD contributions).

Here, we will compute one-loop and (gluon) radiative contributions through
the order $\alpha_{\rm S}^2\alpha_{\rm{W}}$, which -- in the case
of quark-antiquark induced subprocesses -- are represented 
schematically by the 
following diagrams:
\vspace*{0.75truecm}
\begin{eqnarray}
\label{as2aW-qq} \nonumber
\begin{picture}(300,-300)
\SetScale{1.0}
\SetWidth{1.2}
\SetOffset(0,-60)
\ArrowLine(30,90)(45,75)
\ArrowLine(45,75)(45,60)
\ArrowLine(45,60)(30,45)
\Gluon(45,75)(60,75){3}{3}
\Text(25,95)[]{\small $q$}
\Text(25,40)[]{\small $\bar q$}
\ArrowLine(60,75)(75,90)
\ArrowLine(60,60)(60,75)
\ArrowLine(75,45)(60,60)
\Gluon(45,60)(60,60){3}{3}
\Text(80,95)[]{\small $b$}
\Text(80,40)[]{\small $\bar b$}
\Text(105,75)[]{$*~[$}
\ArrowLine(130,90)(145,75)
\ArrowLine(145,75)(130,60)
\Photon(145,75)(160,75){3}{3}
\Text(125,95)[]{\small $q$}
\Text(125,55)[]{\small $\bar q$}
\ArrowLine(160,75)(175,90)
\ArrowLine(175,60)(160,75)
\Text(180,95)[]{\small $b$}
\Text(180,55)[]{\small $\bar b$}
\Text(245,75)[]{$]^\dagger~~ \ +  $ crossed box $+$}
\end{picture} 
\end{eqnarray}
\vspace*{0.75truecm}
\begin{eqnarray}
 \nonumber
\begin{picture}(300,-300)
\SetScale{1.0}
\SetWidth{1.2}
\SetOffset(0,-60)
\ArrowLine(30,90)(45,75)
\ArrowLine(45,75)(45,60)
\ArrowLine(45,60)(30,45)
\Photon(45,75)(60,75){3}{3}
\Text(25,95)[]{\small $q$}
\Text(25,40)[]{\small $\bar q$}
\ArrowLine(60,75)(75,90)
\ArrowLine(60,60)(60,75)
\ArrowLine(75,45)(60,60)
\Gluon(45,60)(60,60){3}{3}
\Text(80,95)[]{\small $b$}
\Text(80,40)[]{\small $\bar b$}
\Text(105,75)[]{$*~[$}
\ArrowLine(130,90)(145,75)
\ArrowLine(145,75)(130,60)
\Gluon(145,75)(160,75){3}{3}
\Text(125,95)[]{\small $q$}
\Text(125,55)[]{\small $\bar q$}
\ArrowLine(160,75)(175,90)
\ArrowLine(175,60)(160,75)
\Text(180,95)[]{\small $b$}
\Text(180,55)[]{\small $\bar b$}
\Text(245,75)[]{$]^\dagger~~\ +$ crossed box $+$}
\end{picture} 
\end{eqnarray}
\vspace*{0.75truecm}
\begin{eqnarray}
 \nonumber
\begin{picture}(300,-300)
\SetScale{1.0}
\SetWidth{1.2}
\SetOffset(0,-60)
\ArrowLine(30,90)(45,75)
\Photon(35,85)(35,65){3}{3}
\ArrowLine(45,75)(30,60)
\Gluon(45,75)(60,75){3}{3}
\Text(25,95)[]{\small $q$}
\Text(25,55)[]{\small $\bar q$}
\ArrowLine(60,75)(75,90)
\ArrowLine(75,60)(60,75)
\Text(80,95)[]{\small $b$}
\Text(80,55)[]{\small $\bar b$}
\Text(105,75)[]{$*~[$}
\ArrowLine(130,90)(145,75)
\ArrowLine(145,75)(130,60)
\Gluon(145,75)(160,75){3}{3}
\Text(125,95)[]{\small $q$}
\Text(125,55)[]{\small $\bar q$}
\ArrowLine(160,75)(175,90)
\ArrowLine(175,60)(160,75)
\Text(180,95)[]{\small $b$}
\Text(180,55)[]{\small $\bar b$}
\Text(265,75)[]{$]^\dagger~~+$ {\normalsize other three vertices~~+}}
\end{picture} 
\end{eqnarray}
\vskip0.5cm
\hskip2.70cm{\normalsize{+~all self-energies~+}}
\vskip-0.05cm
\vspace*{0.75truecm}
\begin{eqnarray}
\begin{picture}(300,-300)
\SetScale{1.0}
\SetWidth{1.2}
\SetOffset(0,-60)
\ArrowLine(30,90)(45,75)
\Gluon(35,85)(45,95){3}{3}
\ArrowLine(45,75)(30,60)
\Gluon(45,75)(60,75){3}{3}
\Text(25,95)[]{\small $q$}
\Text(25,55)[]{\small $\bar q$}
\ArrowLine(60,75)(75,90)
\ArrowLine(75,60)(60,75)
\Text(80,95)[]{\small $b$}
\Text(80,55)[]{\small $\bar b$}
\Text(105,75)[]{$*~[$}
\ArrowLine(130,90)(145,75)
\Gluon(170,85)(155,95){3}{3}
\ArrowLine(145,75)(130,60)
\Photon(145,75)(160,75){3}{3}
\Text(125,95)[]{\small $q$}
\Text(125,55)[]{\small $\bar q$}
\ArrowLine(160,75)(175,90)
\ArrowLine(175,60)(160,75)
\Text(180,95)[]{\small $b$}
\Text(180,55)[]{\small $\bar b$}
\Text(260,75)[]{$]^\dagger~+$ {\normalsize{gluon permutations.}}~~}
\end{picture} 
\end{eqnarray}
The gluon bremsstrahlung graphs are needed in order
to cancel the infinities arising in the virtual contributions when
the intermediate gluon becomes infrared.
 Furthermore, one also has to include 
$\alpha_{\rm S}^2\alpha_{\rm{W}}$ terms induced by gluon-gluon
scattering, that is, interferences between the graphs displayed in
Fig.~1 of Ref.~\cite{EMR} and the tree-level ones for
$gg\to b\bar b$. In the remainder of this paper, we will
assume $m_b=0$ and $m_t=175$ GeV (with $\Gamma_t=1.55$ GeV): the
 top-quark enters the
vertices and self-energies of the diagrams in
(\ref{as2aW-qq}) as well as the boxes (in additions to 
self-energies and vertices themselves)
in Fig.~1 of Ref.~\cite{EMR}, whenever a virtual $W$ exchange occurs. 
The $Z$ mass used was
$M_Z=91.19$ GeV and was related to the $W$ mass, $M_W$, via the
SM formula $M_W=M_Z\cos\theta_W$, where $\sin^2\theta_W=0.232$.
(Corresponding widths were $\Gamma_Z=2.5$ GeV and $\Gamma_W=2.08$ GeV.)
For $\alpha_{\rm S}$ we have used the one- or two-loop expressions
as specified below, with $\lms$ set according to the PDFs used.

Some of the diagrams contain ultraviolet divergences.
These have been subtracted using the `modified' Minimal Subtraction
($\MSbar$) scheme at
the scale $\mu=M_Z$. Thus the couplings are taken to be
those relevant for such a subtraction: e.g., the EM coupling,
$\alpha_{\mathrm{EM}}\equiv \alpha_{\mathrm{EW}}\sin^2\theta_W$, 
has been taken to be $1/128$ at the above subtraction
point. The one exception to this
renormalisation scheme has been the case of the self-energy insertions
on external fermion lines, which have been subtracted on mass-shell,
so that the external fermion fields create or destroy particle states
with the correct normalisation.

Infrared divergences  occur when the virtual or real (bremsstrahlung)
gluon is either soft or collinear with the emitting parton. 
It is because we are considering $b$-jets 
which include a possible gluon parallel to the $b$-quark
rather than open $b$-quark production 
that the collinear divergences cancel, this way removing the
logarithmic dependence on the $b$-quark mass which was investigated and
resummed in the analysis of Ref.~\cite{cgn}.
Moreover, in our case the  collinear
divergences cancel amongst themselves. This can be seen since by colour
conservation only interferences between gluon emission from the initial
and final state quarks are permitted. If the  gluon is parallel
to an initial (final) quark then from  the collinear vertex 
it is contracted into its own momentum and the sum of amplitudes
for a longitudinal gluon emitted from both final (initial) states
cancels by virtue of a Ward identity. For virtual corrections, the 
infrared divergences arise from the box graphs and there is an equivalent 
cancellation of collinear divergences between the crossed and uncrossed
boxes. This leaves the soft divergences which can be readily extracted
 and as expected cancel between the virtual corrections and bremsstrahlung
emissions. Nevertheless, for the sake of numerical stability
when carrying out the necessary  numerical integration over phase space
and convolution with the PDFs, it is preferable to use the formalism
of Catani and Seymour \cite{CS}, whereby corresponding dipole terms 
are subtracted from the bremsstrahlung contributions in order to render the
 phase space integral free of infrared divergences. The
 integration over the gluon phase-space of these dipole terms
 are performed analytically in $d-$dimensions, yielding pole terms
which cancel explicitly against the pole terms of the box graphs.

Our  expressions for each of the diagrams contain the complete
helicity information from both the initial and final state. They
have been calculated using FORM \cite{FORM} and reproduced by an
independent program based on FeynCalc \cite{FeynCalc}. The
formulae have all been checked for gauge  invariance.
The full expressions for the contributions from all possible
$\alpha_{\rm S}^2\alpha_{\rm {EW}}$ graphs are too
lengthy to be reproduced here.

\section*{Numerical results for Tevatron and LHC}

\begin{figure}[!ht]
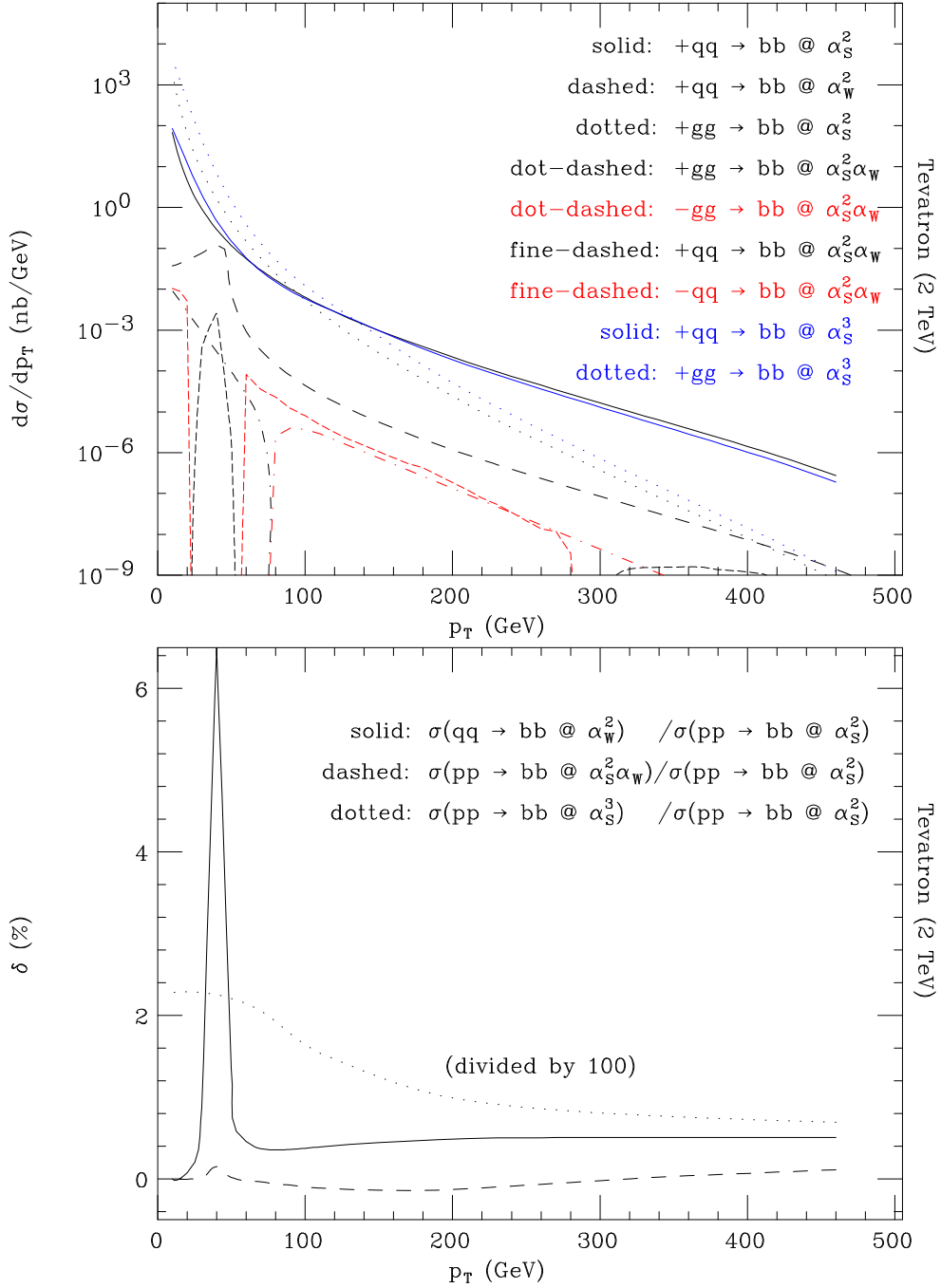

\begin{center}
\vspace{-0.5cm}
{\epsfig{file=sigmabb_Tev.ps, width=9cm, angle=90}}
{\epsfig{file=ratiobb_Tev.ps, width=9cm, angle=90}}
\end{center}
\vspace*{-0.75cm}
\caption{The total cross section contributions vs. the transverse momentum
of the $b$-jet for $p\bar p\to b\bar b$ production at
Tevatron (2 TeV) as obtained via the various subprocesses discussed in the
text (top) and the corrections due to the 
$\alpha_{\rm{EW}}^2$, $\alpha_{\rm{S}}^2\alpha_{\rm{EW}}^2$ 
and $\alpha_{\rm{S}}^3$ terms
relative to the $\alpha_{\rm{S}}^2$ ones (bottom).}
\label{sigmabb_Tev}
\end{figure}

We start our numerical investigation of the processes $pp/p\bar p\to b\bar b$
by first computing the total cross section,  $\sigma(p\bar p\to b\bar b)$,
for Tevatron (Run 2). This can be found in Fig.~\ref{sigmabb_Tev} (top), 
as a function of the transverse momentum of the $b$-jet (or
$\bar{b}$-jet) and decomposed in terms of 
the various subprocesses discussed so far.  (Hereafter, 
the pseudorapidity is limited between $-2$ and $2$ in the partonic CM frame.)
The dominance at inclusive
level of the pure QCD contributions is manifest, over the entire
$p_T$ spectrum. At low transverse momentum it is the gluon-gluon
induced subprocess that dominates, with the quark-antiquark one
becoming the strongest one at large $p_T$. The QCD $K$-factors, defined
as the ratio of the $\alpha_{\rm{S}}^3$ rates to the $\alpha_{\rm{S}}^2$ 
ones are rather large, of order 2 and positive for the $gg\to b\bar b$
subprocess and somewhat smaller  for
the $q\bar q\to b\bar b$ case, which has a $p_T$-dependent
sign\footnote{Further notice that
in QCD at NLO one also has (anti)quark-gluon induced (tree-level) 
contributions, which are of similar strength to those via 
gluon-gluon and quark-antiquark scattering but which have not been shown 
here.}. The tree-level $\alpha_{\rm{EW}}^2$ terms are much smaller
than the QCD rates, typically by three orders of magnitude, with the
exception of the $p_T\approx M_Z/2$ region, where one can appreciate
the onset of the $Z$ resonance in $s$-channel. All above terms are
positive. The $\alpha_{\rm{S}}^2\alpha_{\rm{EW}}$ subprocesses display
a more complicated structure, as their sign can change over the
transverse momentum spectrum considered, and the behaviour is different
in $q\bar q\to b\bar b(g)$ from $gg\to b\bar b$.
Overall, the rates for the $\alpha_{\rm{S}}^2\alpha_{\rm{EW}}$ 
channels are smaller by a factor
of four or so, compared to the tree-level $\alpha_{\rm{EW}}^2$
cross sections.  
Fig.~\ref{sigmabb_Tev} (bottom) shows the percentage
contributions of the   $\alpha_{\rm{S}}^3$, 
$\alpha_{\rm{EW}}^2$ and
$\alpha_{\rm{S}}^2\alpha_{\rm{EW}}$ subprocesses, with respect
to the leading $\alpha_{\rm{S}}^2$ ones, defined as the ratio of each of 
the former to the latter\footnote{In the case of
the $\alpha_{\rm{S}}^3$ corrections, we have used the two-loop 
expression for $\alpha_{\rm{S}}$ and a NLO fit for the
PDFs, as opposed to the one-loop formula and LO set for the 
other processes  (we adopted the GRV94 \cite{PDFs} 
PDFs with $\MSbar$ parameterisation).}.  
The $\alpha_{\rm{S}}^2\alpha_{\rm{EW}}$ 
terms represent a correction of the order
of the fraction of percent to the leading $\alpha_{\rm{S}}^2$ terms.
Clearly, at inclusive level, the effects of the Sudakov logarithms
are not large at Tevatron, 
this being mainly due to the fact that in the partonic
scattering processes the hard  scale involved is not much larger
than the $W$ and $Z$ masses.

\begin{figure}[!ht]
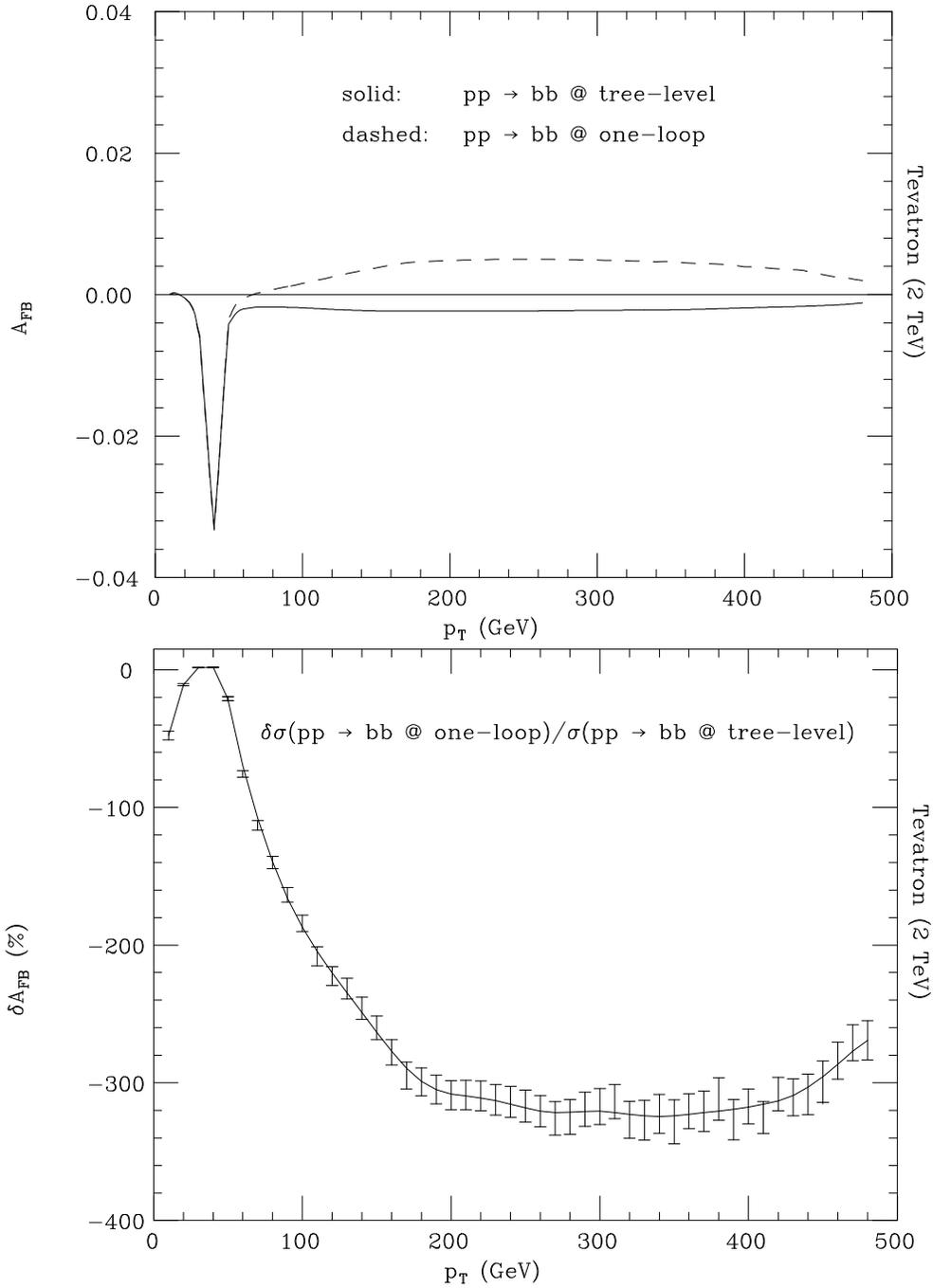

\begin{center}
\vspace{-0.5cm}
{\epsfig{file=sigmabb_AFB_Tev.ps, width=9cm, angle=90}}
{\epsfig{file=ratiobb_AFB_Tev.ps, width=9cm, angle=90}}
\end{center}
\vspace*{-0.75cm}
\caption{The forward-backward  
asymmetry vs. the transverse momentum
of the $b$-jet for $p\bar p\to b\bar b$ events at
Tevatron (2 TeV), as obtained at tree-level and one-loop order (top)
and the relative correction of the latter to the former (bottom).
(Errors in the ratio are statistical.)}
\label{sigmabb_AFB_Tev}
\end{figure}

Next, we study the above mentioned forward-backward asymmetry, defined as
follows:
\begin{equation}\label{AFB}
A_{\rm{FB}}=
\frac{\sigma_+(p\bar p\to b\bar b)-\sigma_-(p\bar p\to b\bar b)}
     {\sigma_+(p\bar p\to b\bar b)+\sigma_-(p\bar p\to b\bar b)},
\end{equation}
where the subscript $+(-)$ iden\-ti\-fies events in which the $b$-jet
is produced with polar angle larger(smal\-ler) than 90 degrees respect to 
one of the two beam directions (hereafter, we use
the proton beam as positive $z$-axis).
The polar angle is defined in the 
CM frame of the hard partonic scattering. Notice that
we do not implement a jet algorithm, as we integrate over the entire phase
space available to the gluon. In practice, this corresponds to 
summing 
over the two- and three-jet contributions that one would extract from the
application of a jet definition. The solid curve in 
Fig.~\ref{sigmabb_AFB_Tev} (top) represents the sum of
 the tree-level contributions
only, that is, those of order $\alpha_{\rm{S}}^2$ and  
$\alpha_{\rm{EW}}^2$, whereas the dashed one also includes the
higher-order ones $\alpha_{\rm{S}}^3$ and
$\alpha_{\rm{S}}^2\alpha_{\rm{EW}}$. (Recall
that the contributions to the asymmetry due to the pure QED and QCD
terms $\alpha_{\rm{EM}}^2$, $\alpha_{\rm{S}}^2$ and $\alpha_{\rm{S}}^3$
are negligible\footnote{And so would also be the one-loop 
$\alpha_{\rm{S}}^2\alpha_{\rm{EM}}$ terms not computed here.}.)

The effects of the one-loop weak corrections on this observable
are extremely large, 
as they are not only competitive with, if not larger than,
the tree-level weak contributions,
but also of opposite sign over most of the considered $p_T$ spectrum.
In absolute terms, the asymmetry is of order $-4\%$ 
at the $W$, $Z$ resonance and
fractions of percent elsewhere, hence it should comfortably be measurable
after the end of Run 2.

\begin{figure}[!ht]
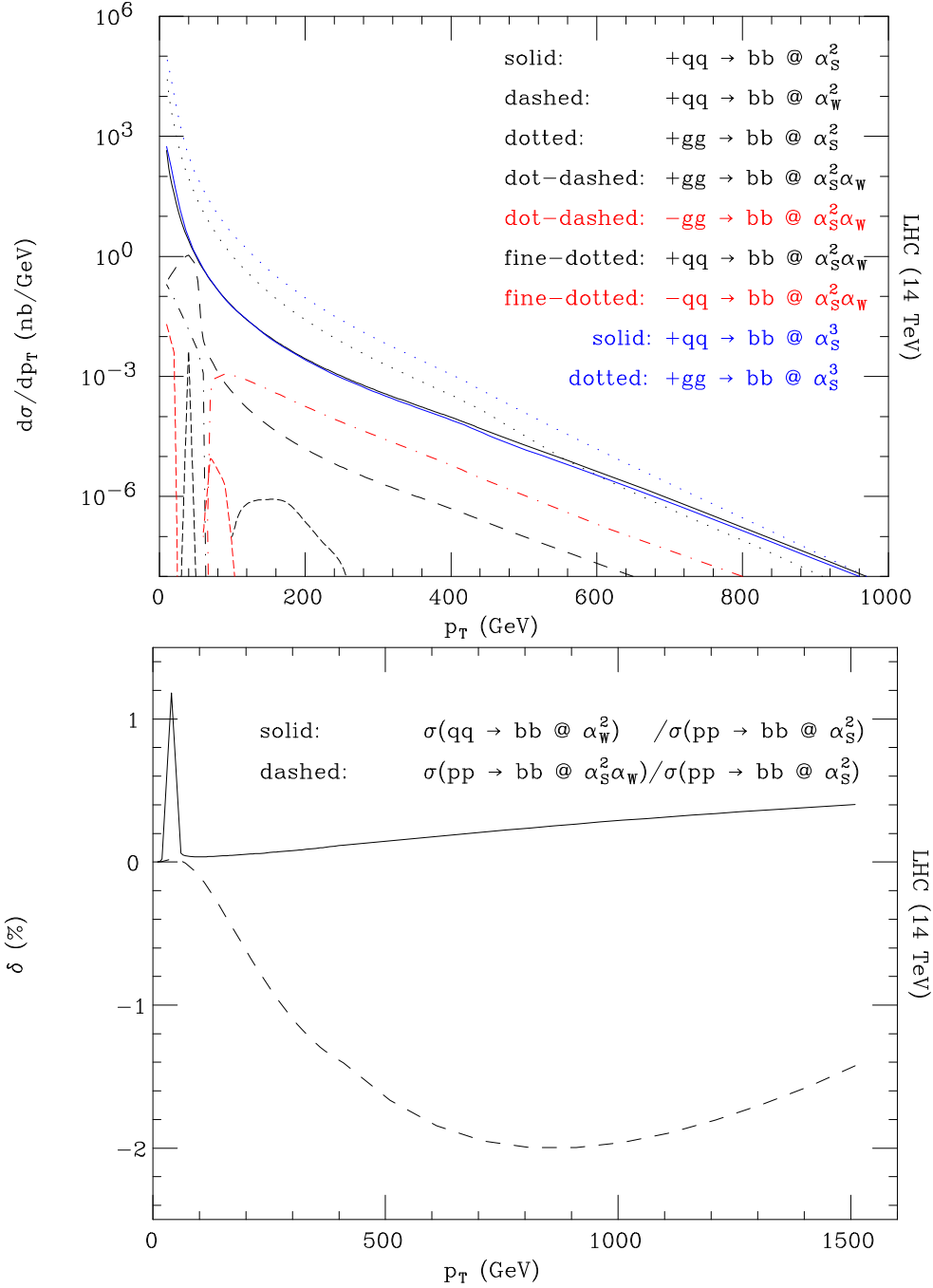

\begin{center}
\vspace{-0.5cm}
{\epsfig{file=sigmabb_LHC.ps, width=9cm, angle=90}}
{\epsfig{file=ratiobb_LHC.ps, width=9cm, angle=90}}
\end{center}
\vspace*{-0.75cm}
\caption{The total cross section contributions vs. the transverse momentum
of the $b$-jet for $pp\to b\bar b$ production at 
LHC (14 TeV) as obtained via the various subprocesses discussed in the
text (top) and the corrections due to the 
$\alpha_{\rm{EW}}^2$ and $\alpha_{\rm{S}}^2\alpha_{\rm{EW}}^2$ terms
relative to the $\alpha_{\rm{S}}^2$ ones (bottom). (Here,
we do not show the corrections due to $\alpha_{\rm{S}}^3$ terms as results are
perturbatively unreliable, given that $K$-factors as large as 3--4 can appear.)}
\label{sigmabb_LHC}
\end{figure}

Fig.~\ref{sigmabb_LHC} shows the same quantities as in 
Fig.~\ref{sigmabb_Tev}, now defined
at LHC energy. By a comparative reading, one may appreciate the 
following aspects. Firstly, the effects at LHC of the 
$\alpha_{\rm{S}}^2\alpha_{\rm{EW}}$ corrections are much larger
than the $\alpha_{\rm{EW}}^2$ ones already at inclusive level
 (see top of Fig.~\ref{sigmabb_LHC}), as
their absolute magnitude becomes of order $-2\%$ or so at large transverse
momentum (see bottom of Fig.~\ref{sigmabb_LHC}): 
clearly, logarithmic enhancements are at LHC much more 
effective than at Tevatron energy scales\footnote{Further notice at LHC
the dominance of the $gg$-induced one-loop terms, as compared
to the corresponding $q\bar q$-ones (top of Fig.~\ref{sigmabb_LHC}), 
contrary to the case of Tevatron, where they
were of similar strength (top of Fig.~\ref{sigmabb_Tev}).}. 
Secondly, the overall production
rates at the CERN collider are in general much larger than those
at FNAL, because of the much larger gluon component of the
proton. 
\section*{Conclusions}

In summary, we should like to remark upon the following aspects of
our analysis. 
\begin{itemize}
\item Inclusive corrections to the $b\bar b$ cross section due to 
one-loop weak interaction
contributions through order $\alpha_{\rm{S}}^2\alpha_{\rm{EW}}$ 
are small and undetectable at Tevatron, while 
becoming visible at LHC, because of the much  larger cross section
and luminosity available. In practice, the weak Sudakov logarithms are
threshold suppressed at the FNAL collider while at the CERN machine
they become sizable.
In the former case then, they cannot explain the current data vs. 
theory discrepancy seen in the  $b$-quark/jet   cross sections. 
\item One-loop weak effects 
onto $b$-quark asymmetries (e.g., we have studied the forward-backward
one) are found to be large at Tevatron, where they  can be 
defined experimentally. Here, the forward-backward asymmetry is 
subject to large corrections 
because the tree-level (quark-antiquark) subprocesses 
are formally of the same order as the one-loop contributions
(initiated by both quark-antiquark and gluon-gluon collisions), eventually
being measurable if collider 
luminosity plans will turn out to be on schedule. 
\end{itemize}
\noindent
In conclusion, at both current and planned TeV scale hadronic
colliders, one-loop weak effects from SM physics may be important and need 
to be taken into account particularly 
in order to extract possible signals of new physics 
from data. 

\section*{Acknowledgments}
MRN, SM and DAR thanks the Department of Theoretical Physics
at Torino University for hospitality while part of this work was being
carried out. SM and DAR are grateful to Stefano Catani and Mike Seymour
for advice. SM thanks Matteo Cacciari for helpful discussions. 
This project was in part financed
by the U.K.\ Particle Physics and
Astronomy Research Council (PPARC), by the 
Italian Ministero dell'Istruzione, dell'Universit\`a e della Ricerca
(MIUR) under
contract 2001023713\_006,
by the European Union under contract HPRN-CT-2000-00149
and by The Royal Society (London, UK)
under the European Science Exchange Programme (ESEP), Grant No. IES-14468.

\newcommand{\plb}[3]{{Phys. Lett.} {\bf B#1}, #3 (#2)}                  %
\newcommand{\prl}[3]{Phys. Rev. Lett. {\bf #1}, #3 (#2) }        %
\newcommand{\rmp}[3]{Rev. Mod.  Phys. {\bf #1}, #3 (#2)}             %
\newcommand{\prep}[3]{Phys. Rep. {\bf #1}, #3 (#2)}                   %
\newcommand{\rpp}[3]{Rep. Prog. Phys. {\bf #1}, #3 (#2)}             %
\newcommand{\prd}[3]{Phys. Rev. {\bf D#1}, #3 (#2)}                    %
\newcommand{\npb}[3]{Nucl. Phys. {\bf B#1}, #3 (#2)}                     %
\newcommand{\npbps}[3]{Nucl. Phys. B (Proc. Suppl.)
           {\bf #1}, #3 (#2)}                                           %
\newcommand{\sci}[3]{Science {\bf #1}, #3 (#2)}                 %
\newcommand{\zpc}[3]{Z.~Phys. C{\bf#1}, #3 (#2)}  
\newcommand{\epjc}[3]{Eur. Phys. J. {\bf C#1}, #3 (#2)} 
\newcommand{\mpla}[3]{Mod. Phys. Lett. {\bf A#1}, #3 (#2)}             %
 \newcommand{\apj}[3]{ Astrophys. J.\/ {\bf #1}, #3 (#2)}       %
\newcommand{\jhep}[3]{{J. High Energy Phys.\/} {\bf #1}, #3 (#2) }%
\newcommand{\jpg}[3]{{J. Phys.\/} {\bf G#1}, #3 (#2)}%
\newcommand{\astropp}[3]{Astropart. Phys. {\bf #1}, #3 (#2)}            %
\newcommand{\ib}[3]{{ibid.\/} {\bf #1}, #3 (#2)}                    %
\newcommand{\nat}[3]{Nature (London) {\bf #1}, #3 (#2)}         %
 \newcommand{\app}[3]{{ Acta Phys. Polon.   B\/}{\bf #1}, #3 (#2)}%
\newcommand{\nuovocim}[3]{Nuovo Cim. {\bf C#1}, #3 (#2)}         %
\newcommand{\yadfiz}[4]{Yad. Fiz. {\bf #1}, #3 (#2);             %
Sov. J. Nucl.  Phys. {\bf #1} #2 (#4)]}               %
\newcommand{\jetp}[6]{{Zh. Eksp. Teor. Fiz.\/} {\bf #1} (#2), #3;
           {JETP } {\bf #4} (#6) #5}%
\newcommand{\philt}[3]{Phil. Trans. Roy. Soc. London A {\bf #1}, #3
        (#2)}                                                          %
\newcommand{\hepph}[1]{hep--ph/#1}           %
\newcommand{\hepex}[1]{hep--ex/#1}           %
\newcommand{\astro}[1]{(astro--ph/#1)}         %

\newpage

\newpage

\end{document}